# Asymptotic Precision Corrections to Radiation Reaction

**Yarden Sheffer[1], Yaron Hadad[1], Morgan H. Lynch[1], Liang Jie Wong[2], Ido Kaminer[1]**

[1] Faculty of Electrical and Computer Engineering, Technion, Israel Institute of Technology, 32000 Haifa, Israel

[2] School of Electrical and Electronic Engineering, Nanyang Technological University, 639798 Singapore, Singapore

kaminer@technion.ac.il

**The radiative correction to the equation of motion for a moving charged particle is one of the oldest open problems in physics. The problem originates in the emission of radiation by an accelerated charge, which must result in a loss of energy and recoil of the charge, adding a correction to the well-known Lorentz force. When radiation reaction is neglected, it is well known that the dynamics of a charge in an ideal plane-wave field is periodic. Here we investigate the long-time dynamics of a charge in such a field and show that all current models of radiation reaction strictly forbid periodic dynamics. Consequently, we show that under the influence of the external field, the loss of energy to radiation reaction causes particles to *accelerate* toward an infinite kinetic energy. Such a phenomenon persists even in weak laser fields and puts forward the possibility of testing radiation reaction through long-duration weak-field precision measurements, rather than through strong-field experiments. We further provide numerical examples suggesting realistic conditions for such measurements through the asymptotic frequency shift and energy loss of a charge, which for example can be detected using electron energy spectrometers in ultrafast electron microscopes.**

## Section I - Introduction

The conventional description for the motion of a charged particle under an electromagnetic field is given by the Lorentz Force (LF) equation of motion [1]. This equation lies at the heart of classical electrodynamics and is known to give an exact description of a wide range of observed physical phenomena. In relativistic covariant form, the LF equation can be written as

$$m\dot{u}^{\mu} = qF^{\mu}_{\ \nu}u^{\nu}. \tag{1}$$

Here $u^{\mu} = \gamma(1, \boldsymbol{\beta})$ is the velocity 4-vector with $\boldsymbol{\beta} = \boldsymbol{v}/c$, $\gamma = 1/\sqrt{1-\beta^2}$ is the Lorentz factor, $F^{\mu\nu}$ is the electromagnetic field tensor, $q$ is the charge and the dot denotes a derivative with respect to the proper time $\tau$. We use $c = 1$ and the metric tensor $g_{\mu\nu} = \{1, -1, -1, -1\}$ throughout.

A well-known result of Maxwell's equations is that an accelerating charged particle emits radiation, with the instantaneous radiation power given by the Larmor formula (here presented for an electron) [1]

$$P = \frac{2}{3}\frac{e^2}{4\pi\varepsilon_0 c^3}\dot{u}^{\mu}\dot{u}_{\mu} = m\tau_0\dot{u}^{\mu}\dot{u}_{\mu}, \tag{2}$$

where $\tau_0 = \frac{2}{3}\frac{e^2}{4\pi\varepsilon_0 mc^3} \approx 6.24 \times 10^{-24}\ [s]$ denotes the characteristic time that it takes light to cross the classical electron radius.

For conservation of energy to be satisfied, radiation emission must be compensated by a loss of energy & momentum from the charge. Consequently, an exact equation of motion for a charged particle must include a radiation correction term, most commonly termed "Radiation Reaction" (RR). The search for the correct treatment of the charged particle dynamics, in a way that consistently include the RR correction, is considered one of the oldest open problems in physics. The current consensus is that a complete treatment for the problem involves quantum electrodynamics, yet the complete recipe for such a treatment is not known. It is therefore believed to be of substantial value to pursue classical and semi-classical treatments of the problem to complement quantum electrodynamics. Such treatments could help guide experiments and provide more insight into the problem of RR.

The most well-known RR correction is the one suggested by Lorentz, Abraham, and Dirac (LAD) [2]:

$$m\dot{u}^{\mu} = qF^{\mu\nu}u_{\nu} + m\tau_0(\ddot{u}^{\mu} + \dot{u}^2 u^{\mu}) \tag{3}$$

Due to the third time derivative of the four-position $x^{\alpha}(\tau)$, the LAD equation does not have a unique solution for a given initial position and momentum [2]. Furthermore, the equation has pathological (runaway) solutions, in which the particle accelerates to infinity, even without the presence of an external field [3]. A proposed remedy was introduced by Dirac, requiring $\dot{u} = 0$ as $t \to \infty$ (this condition, however, replaces the problem of diverging solution with the problem of pre-accelerating solutions, see [4]). Other models of RR were also introduced, prominent examples include Eliezer [5], Landau and Lifshitz [6], Mo and Papas [7], Hartemann and Luhmann [8] and others [9, 10, 11].

Among the various alternative equations, a particularly well-regarded one was introduced by Landau and Lifshitz [6]. By taking the RR correction term in the LAD equation as a perturbation on the Lorentz force (via the minimal substitution $\dot{u}^{\mu} = (q/m)\, F^{\mu\nu}u_{\nu}$), one gets the Landau-Lifshitz (LL) equation:

$$m\dot{u}^{\mu} = qF^{\mu\nu}u_{\nu} + q\tau_0\left\{F^{\mu\nu}_{,\eta}u_{\nu}u^{\eta} + \frac{q}{m}\left[F^{\mu\nu}F_{\nu\eta}u^{\eta} - F^{\nu\eta}F_{\eta\rho}u^{\rho}u_{\nu}u^{\mu}\right]\right\}, \tag{4}$$

This equation is identical to equation (3) to first order in $\tau_0$. The LL equation is often taken for further study, as it avoids the runaway solutions of the LAD equation and has known analytical solutions [12, 13, 14, 15].

Because of the current lack of experimental data, none of the suggested RR models is unanimously accepted, even if just as a classical approximation to the full quantum dynamics. In recent years, however, advancements in high-intensity laser experiments enabled a first empirical look into this old problem [16, 17]. Both experiments reported evidence of significant quantum effects in RR in the measured regimes (field intensities of $10^{18}\ \mathrm{Watt/cm^2}$ and above). Another recent experiment showed the effects of quantum RR for channeled positrons in silicon crystals [18]. As of today, no theory was shown to fully explain the data, which illustrates the persistence of this century-old paradox.

To our knowledge, nearly all proposals and attempts to measure RR assume extreme an interaction with very high intensity fields (with the notable exception of [19]). In such interactions, the RR force is significant or even becomes dominant. Our work discusses a different approach to aid the investigation and observation of RR, relying on low-intensity lasers and precision measurements. Our goal is to quantify measurable effects that can be accessed in the presence of relatively weak electromagnetic fields, and whose description gives new insight about RR. We find such effects by studying the long-time dynamics of a charge under a plane-wave field.

It is well-known that the solution of the LF equation for a charge in a periodic plane wave is also periodic [20]. The effect of RR is a small perturbation on the LF equation that causes loss of energy and acts as damping, therefore altering the original trajectory. One might then expect the charge to converge into an altered steady state, in which an average energy gain from the laser pump is matched by an energy loss to radiation. We show here that regardless of what RR term is used, there exist no steady-state solutions. Instead, we discuss the intriguing result that including RR effects into a long-enough interaction with a plane-wave field causes the particle to accelerate. Moreover, this acceleration extends as long as the field lasts, toward an infinite kinetic energy at infinite time. This supposed divergence is, of course, limited in practice by the duration of the interaction. Such an acceleration effect at long times can be inferred from different methods of analysis of RR [21, 22, 23, 24].

As a first example, we consider the analytic solution of the LL equation and show it predicts a perpetual increase of the particle's momentum to infinity along the direction of the driving plane wave. That is, a charged particle interacting with the wave will be eventually "carried" by it and accelerated toward the speed of light. Next, we consider cases of a very strong radiation force, in which the LL approximation cannot be used, and the dynamics is instead modeled by the LAD equation. We present a proof that even in that case, the dynamics must always asymptotically diverge toward infinite energy, and in fact, the divergence is the same as the one in the LL equation. Lastly, we develop a perturbative method for calculating the RR correction for a general model and use it to show that, under very general assumptions, RR effects prevent the particle from having a periodic motion. Consequently, in all these cases, the same perpetual acceleration toward the speed of light is expected.

One might ask whether there is a difference between the effect of RR under a weak plane-wave field and the RR effect observed in a synchrotron. Synchrotron RR was studied extensively in the fields of accelerator science and engineering, where it is attributed to the energy loss as calculated using the Larmor formula [25, 26]. Here, in contrast, we observe that RR effects enable us to accelerate the particle, an effect that cannot be accounted for by a simple application of the Larmor formula and requires an introduction of a radiation term in the equations of motion. While the two phenomena are governed by similar equations of motion, the resulting dynamics are different, due to the fact that the plane-wave field acts as an energy source. We further note that the acceleration we discuss here is different from other mechanisms of acceleration due to electron-field interactions, for example, ponderomotive acceleration [20], or acceleration due to coherent interaction of electron bunches [27].

After discussing the mathematical side of the acceleration as a result of RR, we conclude the paper by proposing a novel method of measuring the RR correction using the newly developed experimental method of ultrafast electron microscopy. We provide numerical examples in which effects of RR, as modeled by classical equations of motion, can be detected. The reason we find such an experimental proposal exciting is that it relies on much weaker laser intensities than those utilized in previous RR experiments, and instead employ the precision-measurement capabilities of electron microscopy to compensate for the weaker interaction. Weaker laser intensities make it possible to use much more stable laser sources (e.g., as used in laser interferometers, frequency combs, and phase-locked femtosecond lasers), which enable implementing prolonged interactions as discussed below.

Light emission by electrons is intrinsically a quantum electrodynamical (QED) process. Our classical electrodynamics analysis can provide a benchmark for a QED analysis of RR [28, 29, 30, 31, 32] and possible future experiments. We expect our prediction of the breaking of plane-wave periodic dynamics to be a feature of the QED description as well, at least in some parameter regimes. The more general question as to whether the breaking of periodic dynamics would always occur for every choice of parameters in QED given a long-enough interaction with a plane-wave field is important and will be left for future work.

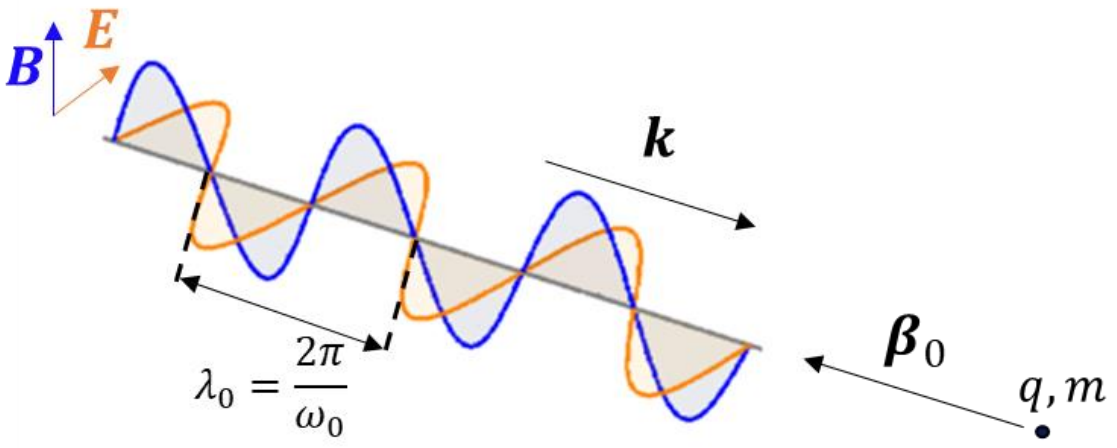

Figure 1: **Outline of the interaction.** Here $\boldsymbol{k}$ is the wave vector (with wavelength $\lambda_0$ and frequency $\omega_0$), and $\boldsymbol{\beta}_0$ is the initial particle velocity.

## Section II - Solution to Lorentz force (LF) equation in a plane wave

As a preliminary to our discussion, we first consider the known solution for the LF equation (1) in the field of a plane wave, written as

$$F^{\alpha}{}_{\beta} = \begin{bmatrix} 0 & 1 & 0 & 0 \\ 1 & 0 & 0 & -1 \\ 0 & 0 & 0 & 0 \\ 0 & 1 & 0 & 0 \end{bmatrix} E_0 \varphi(k \cdot x), \tag{5}$$

where $\varphi(k \cdot x)$ represents an arbitrary pulse envelope, with the dot product defined as $a \cdot b = a^{\alpha} b_{\alpha}$, and $k^{\alpha} \propto (1,0,0,1)$ is chosen for simplicity to describe propagation in the $+z$ direction with linear polarization in the $x$ direction (it is straightforward to generalize it to other cases). Specifically, later in this work, we use a monochromatic plane wave with a frequency $\omega_0$ so that $k^{\alpha} = (\omega_0, 0,0, \omega_0)$ and $\varphi(k \cdot x)$ is a sine wave. An important property of $F^{\mu\nu}$, which will be used later, is that

$$k_{\mu} F^{\mu\nu} = 0. \tag{6}$$

Due to the plane-wave excitation, the LF equation seems to be nonlinear in $u$ (since $\dot{x} = u$, with $x$ appearing inside $F^{\alpha}{}_{\beta}$). But it can be linearized by introducing the change of variables $\xi = k \cdot x$ (representing the phase of the wave at the point of the particle), which satisfies

$$\ddot{\xi} = k \cdot \dot{u} = 0 \tag{7}$$

so $\dot{\xi}$ is constant and given by

$$\dot{\xi} = k \cdot u_{\text{in}}, \tag{8}$$

where $u_{\text{in}}$ is the initial velocity. Equation (1) then becomes

$$(u^{\nu})' = \frac{q}{m\,(k \cdot u_{\text{in}})} F^{\mu}{}_{\nu} u^{\nu}. \tag{9}$$

Here the prime denotes differentiation with respect to $\xi$.

We can now use the fact that the matrices $F^{\mu}{}_{\nu}(\xi)$ at different times commute with each other to integrate equation (9). We then get

$$u^{\mu}(\xi) = U^{\mu}{}_{\nu}(\xi) u^{\nu}_{\text{in}}\,, \tag{10}$$

where $U^{\mu}{}_{\nu}$ is the propagator

$$U^{\mu}{}_{\nu}(\xi) = \begin{bmatrix} 1 + \frac{\chi^2(\xi)}{2} & \chi(\xi) & 0 & -\frac{\chi^2(\xi)}{2} \\ \chi(\xi) & 1 & 0 & -\chi(\xi) \\ 0 & 0 & 1 & 0 \\ \frac{\chi^2(\xi)}{2} & \chi(\xi) & 0 & 1 - \frac{\chi^2(\xi)}{2} \end{bmatrix} \tag{11}$$

$$\chi(\xi) = \frac{a_0}{\hat{k} \cdot u_{\text{in}}} \int_0^{\xi} \varphi(\tilde{\xi}) d\tilde{\xi}\,, \tag{12}$$

with $a_0 = \frac{E_0 q}{m\omega_0}$ being the dimensionless parameter describing the field intensity, and $\hat{k}^{\mu} = \frac{1}{\omega_0} k^{\mu}$ ($\tilde{\xi}$ will be used throughout as an integration parameter when replacing $\xi$).

An important property of this solution is that for a periodic wave (periodic $\varphi$), the solution $u^{\mu}$ is periodic in time. Furthermore, when $\varphi$ has a zero mean (that is, no DC field), the particle cannot be accelerated, i.e., gain net energy over the entire interaction. This result is generalized as the Lawson-Woodward theorem [33, 34], which states that no first-order acceleration can occur for a charged particle interacting with a laser in free space (under some additional conditions [35]).

**Section III - Perpetual accceleration under the Landau-Lifshitz (LL) equation & the Lorentz-Abraham-Dirac (LAD) equation**

In contrast to the LF equation that shows periodic dynamics, we find that the long-time dynamics of the LL equation has the particle energy increase indefinitely. The LL equation of motion for a particle traveling in a plane wave was solved analytically by di Piazza [14] and by Hadad et al. [15] (here we follow the notations of [15], with the exact solution presented in equation (31) therein). It is worthwhile, however, to give a further discussion of the asymptotic dynamics of the solution. As RR is a perturbation that corresponds to a damping force, one might expect that the solution will decay to a periodic solution. This is not the case, however, and we shall further see that the LL model predicts a perpetual increase of the particle energy to infinity in all cases. To see this, consider the functions (taken from [15])

$$k \cdot u(\xi) = \frac{k \cdot u_{\mathrm{in}}}{1 + \tau_0 a_0^2 (k \cdot u_{\mathrm{in}})\, \psi(\xi)} \tag{13}$$

and

$$\psi(\xi) = \int_0^{\xi} \varphi^2(\tilde{\xi}) d\tilde{\xi}\,, \tag{14}$$

with $\varphi$ defined in equation (5). It can now be seen that when the particle's interaction with the EM pulse is not bounded in time, equation (14) enforces $\psi \to \infty$ and thus according to equation (13) we also have $\gamma(1-\beta_z) = k \cdot u \to 0$. This is possible only if $\beta_z \to 1$ (and because $|\beta| < 1$ also $\beta_x, \beta_y \to 0$). That is, RR leads to a counter-intuitive result: it enables the acceleration of the electron, as was suggested by [36] in the context of bichromatic laser fields, and in [21] in the case of an electron initially at rest. In fact, our argument shows that the laser can pump an infinite amount of energy to the electron. Moreover, an eventual unbounded acceleration occurs for an electron starting at any velocity, and it already appears for a monochromatic laser field.

We note that a similar conclusion can be reached when considering the solution to the acceleration due to the averaged RR force, as was done by Landau & Lifshitz [6, p. 219].

Another important note is that while it might be tempting to associate the perpetual acceleration seen here to the divergent "runaway" solutions of the LAD equation, they are completely distinct. The runway solutions (see, for example, [2, 37]) are singular in $\tau_0$, that is, the solution diverges faster as $\tau_0 \to 0$, while here the divergence is regular in $\tau_0$.

We also identify a frequency shift, i.e., the change in the particle's oscillation frequency resulting from the RR correction to the particle trajectory. The frequency shift can be calculated by first presenting the relation between the time $t$ and the variable $\xi$

$$t(\xi) = \int_0^{\xi} \frac{dt}{d\tau}\frac{d\tau}{d\tilde{\xi}} d\tilde{\xi} = \int_0^{\xi} \dot{t}(\tilde{\xi})\tau'(\tilde{\xi}) d\tilde{\xi} = \int_0^{\xi} \frac{u^0}{k \cdot u} d\tilde{\xi}. \tag{15}$$

When the integrand changes slowly, we can also define

$$\omega(\xi) = \frac{2\pi}{t(\xi + 2\pi) - t(\xi)} = 2\pi \left( \int_{\xi}^{\xi + 2\pi} \frac{u^0}{k \cdot u} d\tilde{\xi} \right)^{-1} \approx \frac{k \cdot u(\xi)}{u^0(\xi)}. \tag{16}$$

Figure 2 shows the long-time dynamics under RR effects in the presence of a relatively weak field. We see that while in a head-on collision of a particle with an EM field, the particle experiences an initial energy loss, eventually the RR effect causes acceleration in the direction of the wave (notice, however, that while the acceleration is unbounded, the acceleration per period is still small).

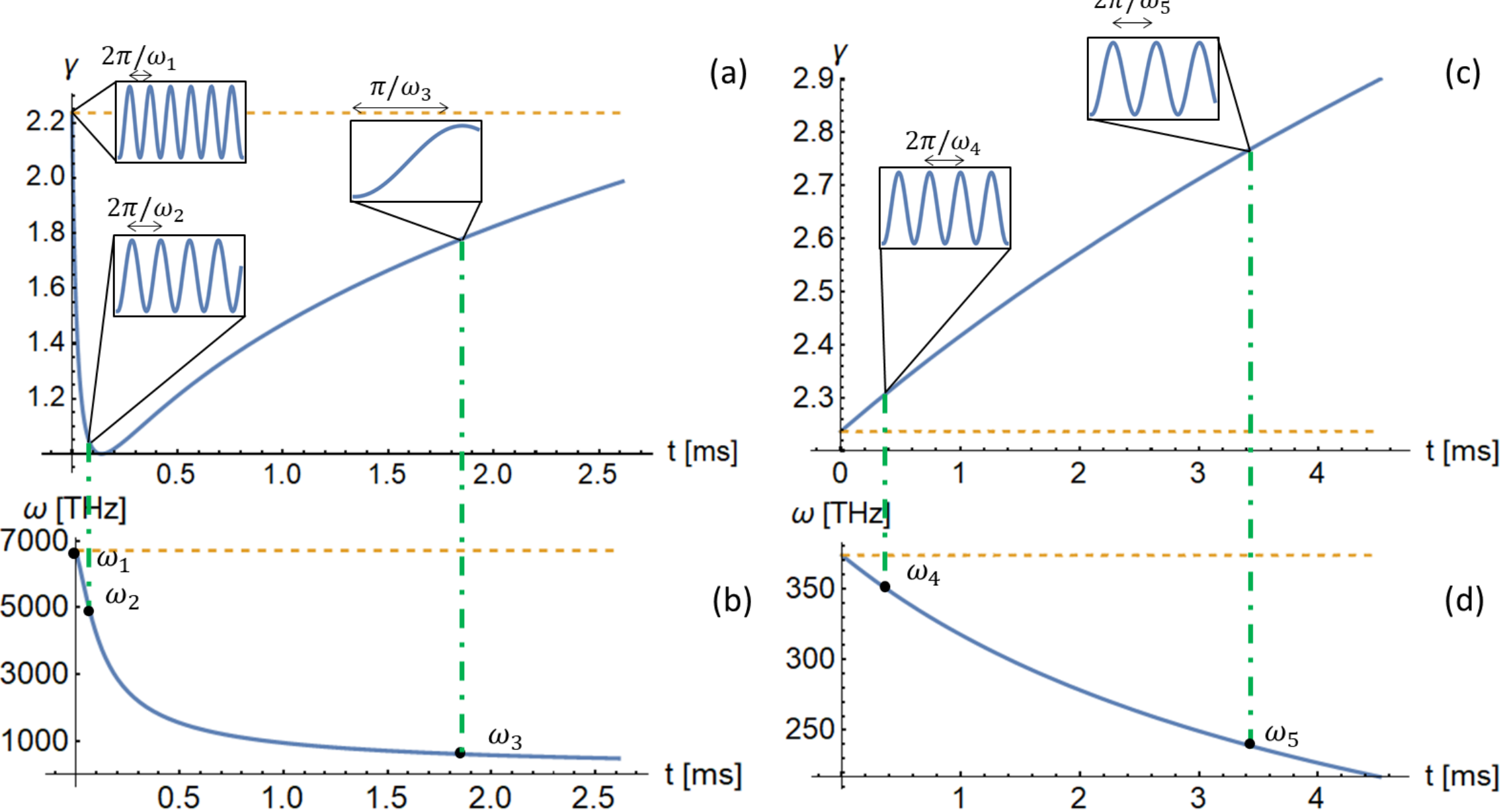

Figure 2: **The long-time dynamics due to RR.** Long time energy (a) and frequency (b) of an electron in the LL equation with initial velocity $u^z = 2$ (kinetic energy of 0.63 MeV) against the direction of the wave vector, and a linearly-polarized plane wave with $a_0 = 0.01$, $\lambda_0 = 532$ nm. In (a), we can see the particle energy decaying at first, as shown in [15], but eventually rising, as a result of the particle long-time acceleration in the direction of the wave vector. Figures (c) and (d) depict the same change in energy and frequency, but for a particle with initial velocity in the direction of the wave.

Since the LL equation is an approximation of LAD, one might wonder whether our result of perpetual acceleration is an outcome of the approximation. For instance, the result will not be valid when the particle energy becomes large enough or when the duration of the interaction becomes long enough. Therefore, it is interesting to ask whether the long-time dynamics governed by the LAD equation can prevent the perpetual acceleration seen in the LL equation, and result in steady-state dynamics. We prove (see Appendix 1) that the LAD equation cannot, in any case, admit a steady-state (periodic) solution for a particle traveling in a plane wave. Our proof predicts that for an arbitrarily strong linearly polarized pulse, the LAD equation *always predicts an eventual increase of energy toward infinity* (resulting from $k \cdot u \to 0$ and eventually converging to the first-order perturbative dynamics described by the LL equation). This is a remarkable result, as the LAD equation is a complex nonlinear system, which in general is expected to show regimes of both periodic and chaotic motion.

### Section IV - Perturbative calculation of general RR corrections

Next, we wish to generalize the solution presented for the LL equation to a general RR term. This can be applied, for example, on a semi-classical RR force obtained by the summation of transitions between quantum Volkov states in a plane wave, as described in [38]. Ultimately, by the end of the next section, we want to prove that no steady-state solution can exist for a broad class of RR models to the first order in $\tau_0$ and consequently that the particle energy must increase indefinitely in long-enough interactions with an EM field. In this section, we develop a perturbation theory for the shift in velocity and energy of the particle by a general covariant RR model. Importantly, the first-order perturbation dominates the effects of RR in the precision measurement experiments that will be described below, where the RR term is indeed a small perturbation on the LF. The first-order phase-space diagrams shown in Figure 3 then provide a sufficient description of the long-term dynamics.

We solve an equation with the general form

$$m\dot{u}^{\mu} = \mathrm{q}F^{\mu}{}_{\nu}u^{\nu} + m\varepsilon D^{\mu}, \tag{17}$$

where $F^{\alpha}{}_{\beta}$ is the EM wave field tensor, $D^{\mu}\big(u, \dot{u}, \dots, F^{\mu\nu}, F^{\mu\nu}{}_{,\rho}, \dots\big)$ is a general RR term (units of $\mathrm{time}^{-2}$), and ε is a small parameter with dimensions of time, satisfying $\varepsilon\omega \ll 1$. We will later add a further scaling assumption to the definition of $D$, while keeping it general enough to include all well-known RR equations [39]. It can be shown (for further details see Appendix 2) that to first order in ε, the 4-velocity of the charge, including the RR effect, can be written as

$$u^{\mu} = U^{\mu}{}_{\nu}u^{\nu}_{\mathrm{in}} + \frac{\varepsilon}{k\cdot u_{\mathrm{in}}}U^{\mu}{}_{\nu}\int_{0}^{\xi}(U^{-1})^{\nu}{}_{\eta}\left[D^{(0)\eta} - X(\tilde{\xi})\frac{d}{d\tilde{\xi}}u^{(0)\eta}\right]d\tilde{\xi} + O(\omega\varepsilon)^{2}, \tag{18}$$

(with terms in the integrand evaluated at $\tilde{\xi}$). $X$ is an auxiliary function

$$X(\xi) = \frac{1}{k\cdot u_{\mathrm{in}}}\int_{0}^{\xi}k\cdot D^{(0)}(\tilde{\xi})d\tilde{\xi}, \tag{19}$$

${U^\mu}_\nu$ being the propagator described in equation (11), and $D^{(0)\mu}$ is $D^\mu$ evaluated on the unperturbed trajectory (all are a function of $\xi$) $D^{(0)\mu} = D^\mu\left(u^{(0)\nu}, \dot{u}^{(0)\nu}, \dots, F^{\nu\eta}, F^{\nu\eta}_{\ ,\rho} \dots\right)$.

We can follow the particle dynamics one cycle of the EM field at a time, by integrating each cycle separately, thus describing the particle dynamics in terms of a discrete-time difference equation. This enables us to analyze the long-time evolution by looking at the difference in velocity $\Delta u^\mu_{\text{in}}$ after each cycle. This difference is only a result of the RR term, as $\Delta u^\mu_{\text{in}}$ is zero in the absence of an RR correction. Consequently

$$\Delta u^\mu_{\text{in}}(u^\mu_{\text{in}}) = \frac{\varepsilon}{k\cdot u_{\text{in}}}\int_0^{2\pi} (U^{-1})^\mu_{\ \nu}\left[D^{(0)\nu} - X(\tilde{\xi})\frac{d}{d\tilde{\xi}}u^{(0)\nu}\right]d\tilde{\xi}\,. \tag{20}$$

Note that $\Delta u^\mu_{\text{in}}$ here is Lorentz invariant, as it is composed of Lorentz-invariant objects.

Figure 3 shows two examples of phase spaces (Poincare sections) for the difference equation (20), defined on the charge velocity $\beta$, and mapping the particle dynamics with the

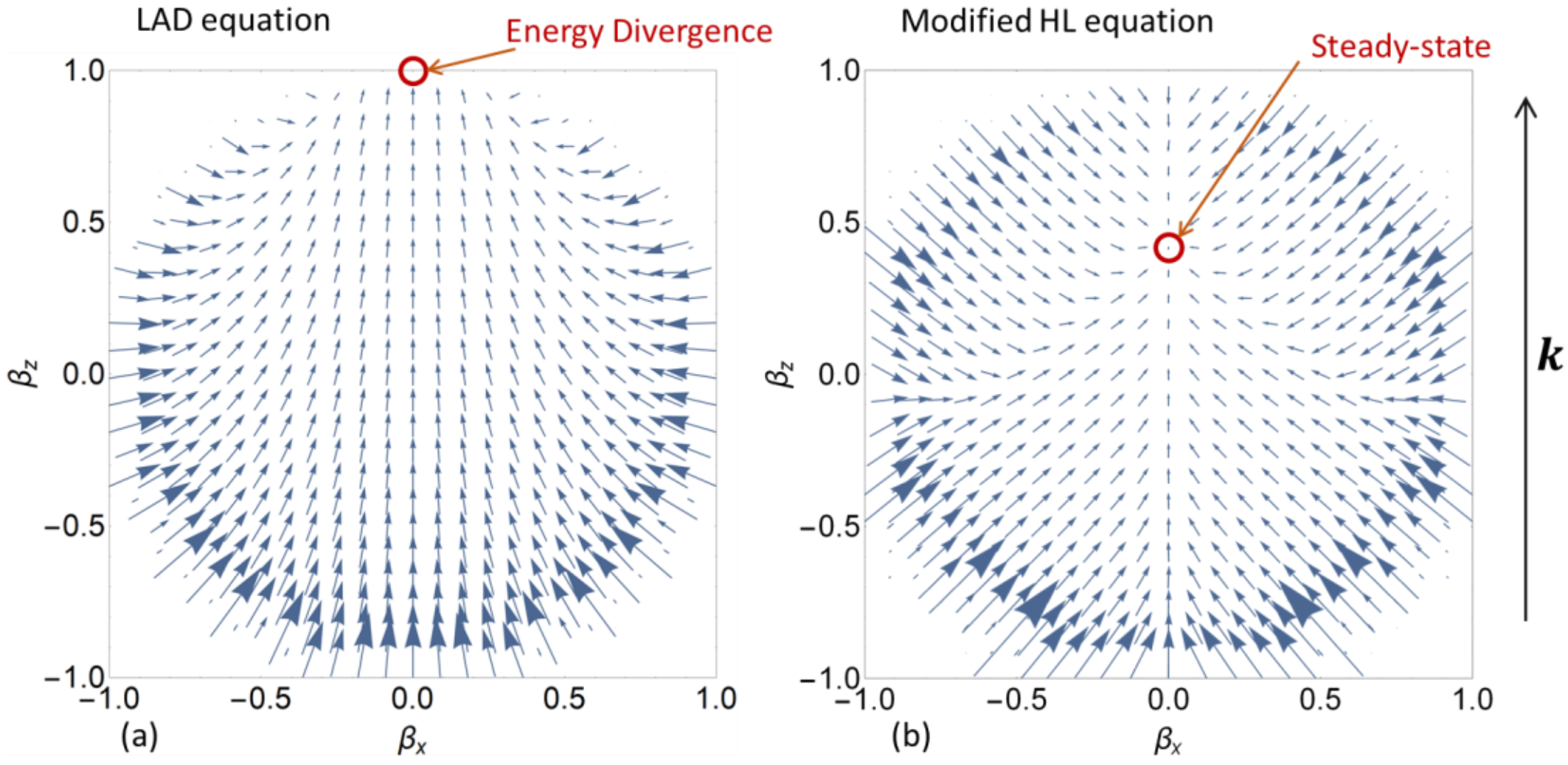

Figure 3: **Calculation of the long-time phase space for the velocity $\boldsymbol{\beta}$ showing energy divergence for the LAD equation**, as calculated using (20) for two different models. An arrows map represents the particle's dynamics: each arrow connects a given $\beta$ to the particle's $\beta$ after a single period of the EM field. The wave vector $\boldsymbol{k}$ points towards the $+z$ direction. (a) Calculated for the LAD equation, exhibiting the same perpetual acceleration as described in Section V. (b) Calculated for a model similar to Hartemann-Luhmann's [8], but with a modified zero term (see Appendix 3). Note that (b) has a steady state point at $\beta_z = 0.4$ (marked in red), and thus cannot be a Lorentz invariant RR model according to our proof.

arrows representing $\Delta u^\mu_{\text{in}}(\beta)$. That is, each arrow in the phase space connects a given velocity

$\beta(\xi)$ to $\beta(\xi + 2\pi)$, its value after a single period of the field. To put it in context, the phase space for the LF equation ($D^\mu \propto F^{\mu\nu}u_\nu$) will be composed of only fixed points (all the arrows are of size zero) since the dynamics is always periodic. This can be shown directly by substituting $D^{(0)\mu} = \text{const} \times F^{(0)\mu\nu}u_\nu^{(0)}$ in equation (20).

**Section V – Proof of the perpetual acceleration for a general RR term**

By analyzing the dynamics in this phase space, we prove in this section that under broad assumptions no Lorentz-invariant equation of motion (i.e., a covariant $D^\alpha$) can reach a periodic solution. Thus, the kinetic energy must always increase toward infinity for a long-enough plane-wave excitation. In the phase space diagram, this means that there exists no fixed point (i.e., $\beta$ that translates to itself) and no cycle (i.e., $\beta$ that translates to itself after $n$ periods), so following the arrows always converges to the edge of the diagram ($|\beta| = 1$). As an example, Figure 3a shows dynamics converging to the edge of the diagram, representing asymptotically diverging dynamics. In contrast, Figure 3b shows an attractive stable point (point with no arrow) in phase space, representing periodic dynamics. Our proof shows that the picture depicted in Figure 3b cannot arise from a covariant RR term.

We prove by contradiction: Assume there exist a fixed $u^\mu_{\text{in},0}$ inside the phase space (finite velocity). This can be thought of as a zero of the function $\Delta u^\mu_{\text{in}}(u^\mu_{\text{in}})$. We prove that $u^\mu_{\text{in},0}$ is unchanged by some Lorentz transformations of the phase space. That is, the transformed function $\Delta u'^\mu_{\text{in}}$ (the change in velocity in the observer frame as a function of initial velocity in the observer frame) has the same zero $u^\mu_{\text{in},0}$. This is in contradiction with the fact that velocities transform as 4-vectors.

To see this, it is sufficient to consider an observer moving in the direction of the wave vector with a velocity $\beta_{\mathrm{ob}}$. This observer will see the interaction of a particle with a field that remains a plane wave and is now characterized by the boosted parameters

$$a_0 \mapsto a_0, \qquad \omega_0 \mapsto \sqrt{\frac{1-\beta_{\mathrm{ob}}}{1+\beta_{\mathrm{ob}}}}\,\omega_0, \qquad F^{\mu\nu}(\xi) \mapsto \sqrt{\frac{1-\beta_{\mathrm{ob}}}{1+\beta_{\mathrm{ob}}}}\,F^{\mu\nu}(\xi). \tag{21}$$

We note that since $F^{\mu\nu}$ generally transforms like a tensor, this transformation is valid only in the case of a plane wave and only when the observer is moving in the direction of the field.

Let us now examine how the arrows in phase space (see Figure 3) $\Delta u_{\mathrm{in}}'^{\mu}(u_{\mathrm{in}}^{\mu})$ are seen by the moving observer (that is, velocity change in the observer frame as a function of some velocity 4-vector $u_{\mathrm{in}}^{\mu}$ in that frame). Since the LF propagator $U^{\mu}{}_{\nu}(\xi)$ is only a function of $a_0$, the transformation of $\Delta u_{\mathrm{in}}'^{\mu}(u_{\mathrm{in}}^{\mu})$ in equation (20) will only be a result of the transformation of $D^{(0)\mu}$. $D^{(0)\mu}$ is a function of $(\frac{d}{d\tau}, \frac{d^2}{d\tau^2}, \dots, F^{\mu\nu}, F^{\mu\nu}_{,\rho} \dots)$ acting on $u^{(0)\mu}$ and thus, noticing that the action of $\frac{d}{d\tau}$ and $F^{\mu\nu}$ on $u^{(0)\mu}(\xi)$ scales as the Doppler factor $\kappa = \sqrt{\frac{1-\beta_{\mathrm{ob}}}{1+\beta_{\mathrm{ob}}}}$, we group the terms in $D'^{(0)\mu}$ by powers of $\kappa$:

$$D'^{(0)\mu}\left(\xi, u_{\mathrm{in}}^{\mu}\right) = \sum_{n} \kappa^n g_n\left(\xi, u_{\mathrm{in}}^{\mu}\right), \tag{22}$$

where $g_n\left(\xi, u_{\mathrm{in}}^{\mu}\right)$ are arbitrary functions of the observer-frame velocity. For example, the terms that are shown in $g_2$ are $\ddot{u}$, $(\dot{u})^2$, $F^{\mu\nu}_{,\rho}$, and so on.

At this point, we use the fact that for a broad class of RR models, including the most commonly discussed ones (LAD, LL, etc.), only the terms corresponding to $n = 2$ are present in the expression, and thus equation (22) simplifies to

$$D'^{(0)\mu}\left(\xi, u_{\mathrm{in}}^{\mu}\right) = \kappa^2 g_2\left(\xi, u_{\mathrm{in}}^{\mu}\right). \tag{23}$$

Substituting in equation (20), we see that it can now be separated as

$$\Delta u'^{\mu}_{\text{in}}\left(u^{\mu}_{\text{in}}\right) = \kappa\, \Delta u^{\mu}_{\text{in}}\left(u^{\mu}_{\text{in}}\right), \tag{24}$$

With $\Delta u^{\mu}_{\text{in}}\left(u^{\mu}_{\text{in}}\right)$ being the velocity change in the lab frame ($\beta_{\text{ob}} = 0$). Notice that in both sides of equation (24), we have the same initial velocity, but it plays the role of velocity in the observer frame on the LHS and velocity in the lab frame on the RHS. The phase space seen by the moving observer will thus have the arrows pointing at the same directions, but with their length scaled by a constant factor $\kappa$. Therefore, if the phase space had a fixed point with an initial velocity $u^{\mu}_{\text{in},0}$ such that $\Delta u^{\mu}_{\text{in}}\left(u^{\mu}_{\text{in},0}\right) = 0$ (as illustrated in Figure 3b), we also have $\Delta u'^{\mu}_{\text{in}}\left(u^{\mu}_{\text{in},0}\right) = 0$. Then, $u^{\mu}_{\text{in},0}$ must also be a steady point in the boosted frame, in contradiction with the requirement that the steady-state velocity must transform as a 4-vector.

The resulting contradiction implies that Lorentz invariance does not allow the existence of periodic points in phase space. We can show that trajectories with longer periods cannot exist either (i.e., the particle returns to its initial state after $n$ periods) by following the same proof, replacing the upper integration limit in equation (20) with $2\pi n$. Dynamics with a period that is not an integer number of cycles are not possible either, as the plane wave EM field is $2\pi$ periodic (in $\xi$).

For the final step in the proof, we show the divergence of the solution toward infinite energy at infinite interaction time. In the limit of weak RR, we can treat the small changes between discrete periods as approximating a continuous flow. In this limit, where the first-order solution is applicable, we have $\frac{\partial \Delta u^{\mu}_{\text{in}}}{\partial u^{\nu}_{\text{in}}} \Delta u^{\nu}_{\text{in}} \ll \Delta u^{\nu}_{\text{in}}$, i.e., the changes in the phase-space flow field are much smaller than the changes of $u^{\mu}$ between periods. Therefore, as the $\beta$ phase space is two-dimensional (in the case of a linearly polarized wave) with no periodic cycles, the divergence of the trajectory is established as a result of the Poincare-Bendixon theorem [40], which states that a 2D continuous dynamical system can either converge into a stable point, a periodic limit cycle, or diverge.

It is instructive to ask what scenarios enable steady-state dynamics. For example, in the case of a particle under the influence of two counter-propagating waves, the field amplitudes of each wave will scale differently upon a Lorentz transform, so that the amplitude of the wave moving counter to the observer increases, and the other decreases. In that case, the total field in different points in space-time scales differently, so that a scaling law similar to the one applied in equation (21) cannot be established, and thus a steady state could potentially be obtained despite the influence of RR.

The inclusion of only the $\kappa^2$ terms in equation (22) is also a fine point that deserves further discussion. This is indeed a property of all well-known RR equations, such as the LAD, LL, Mo & Papas [41] and Sokolov (to the first order) [9] equations, and stems from the scaling of the radiation power due to the Larmor formula as $\dot{u}^2$. However, quantum corrections to the Larmor formula add higher-order terms in frequency (see, for example [42, pp. 376-386]), so that a phenomenological model that accounts for corrections of quantum electrodynamics (QED) should display effects of orders higher than $\left(\frac{d}{d\tau}\right)^2$ (i.e., higher than $\kappa^2$), on which our proof is not applicable [43]. This introduction of higher-order terms leads to a fascinating implication: that convergence to a steady state might be possible, and if found in experiments, hints at a necessarily non-classical behavior that is inherent in RR. We note, however, that the most well-known phenomenological QED model, by Sokolov, exhibits a similar acceleration to the one shown above [44].

## Section VI – Quantitative examples for possible precision measurements of RR

We now discuss the implications of the above findings to possible precision experiments that may give insight into the problem of RR. If the dynamics of a particle under the influence of a plane-wave EM field would be found in some candidate RR equation to have an altered steady-state solution, such a steady state could be measured by trapping the particle and using precision measurement tools to observe corrections to its motion after a long interaction. Interestingly, in the previous sections we showed that, for all the famous RR candidate models, the dynamics of a charged particle interacting with a periodic wave always diverges toward infinite energy. Below, we estimate the size of the RR corrections and discuss possible precision experiments given a long enough interaction.

We consider the effect of the interaction on the shift in the energy of the particle (see an outline in Figure 4). Figure 5 shows the time that it takes a particle to change its energy by 1 eV. We further calculate the energy loss due to the LL correction $\Delta E$, and find the laser pulse fluence $H$ (i.e., the total energy per unit area) required to obtain this energy loss. The relation is captured by the simple equation (see Appendix 4 for derivation)

$$H = \frac{c m_e \varepsilon_0 \Delta E}{q^2 \tau_0}\left(1 + \frac{1}{\beta_{\mathrm{in}}}\right) = 242[\mathrm{kJ/cm^2}] \times \left(1 + \frac{1}{\beta_{\mathrm{in}}}\right)\frac{\Delta E}{1\ \mathrm{eV}}\ . \tag{25}$$

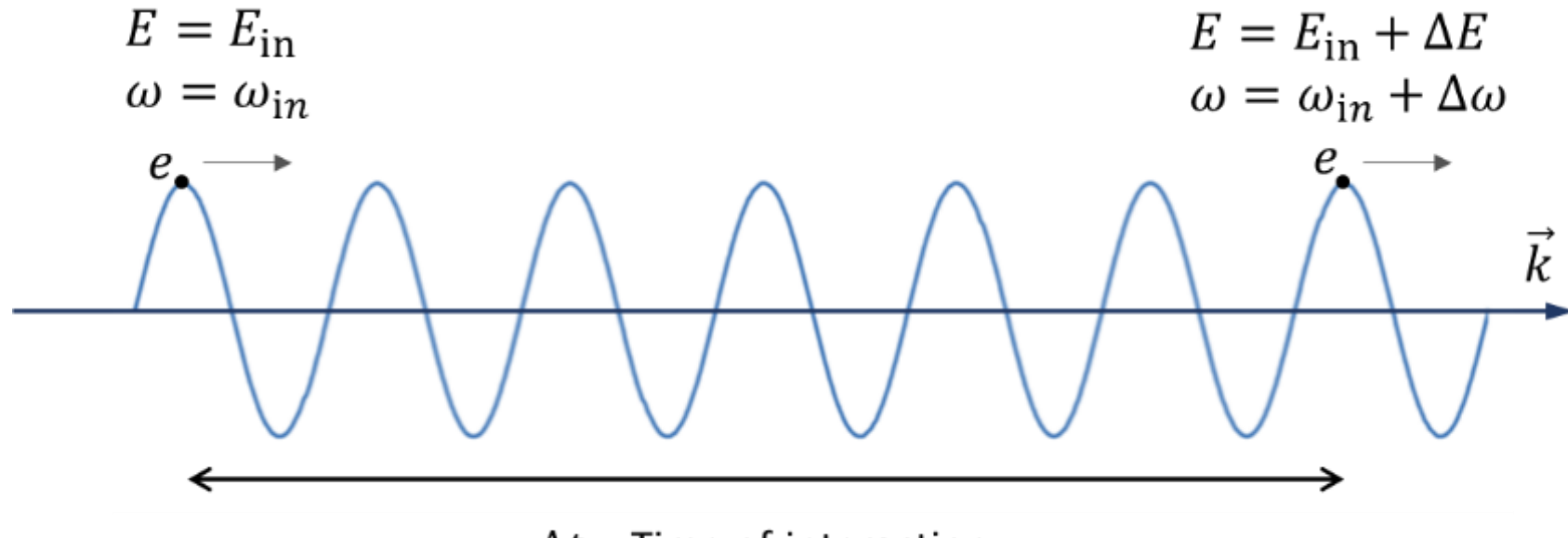

Figure 4: **Energy & Frequency shift of the electron.**

where $\Delta E$ is assumed to satisfy $\left|\frac{\Delta E}{\beta mc^2}\right| \ll 1$. The relation above assumes a linearly polarized plane wave but is otherwise independent of the specific laser parameters: frequency, intensity, pulse envelope and duration.

An especially attractive opportunity for such an experiment is the use of ultrafast/dynamic transmission electron microscopes (UTEM/DTEM) [45, 46, 47] or ultrafast electron diffraction (UED) [48, 49]. Such setups facilitate the interaction of fs/ns laser pulses with free electrons. Using electron energy spectrometers, such systems allow the precise measurement of the electron energy/momentum change, with precisions on the scales of a single eV and down to a few meV in state-of-the-art systems [50, 51].

Another intriguing opportunity presented by such precision experiments is to conduct the interaction inside a hollow optical waveguide [52, 53]. Such waveguides will enable longer interaction lengths with stronger laser fields. The typical interaction length is then bounded by the electron beam spread angle (which can be kept much below $1\ \mathrm{mrad}$) instead of the

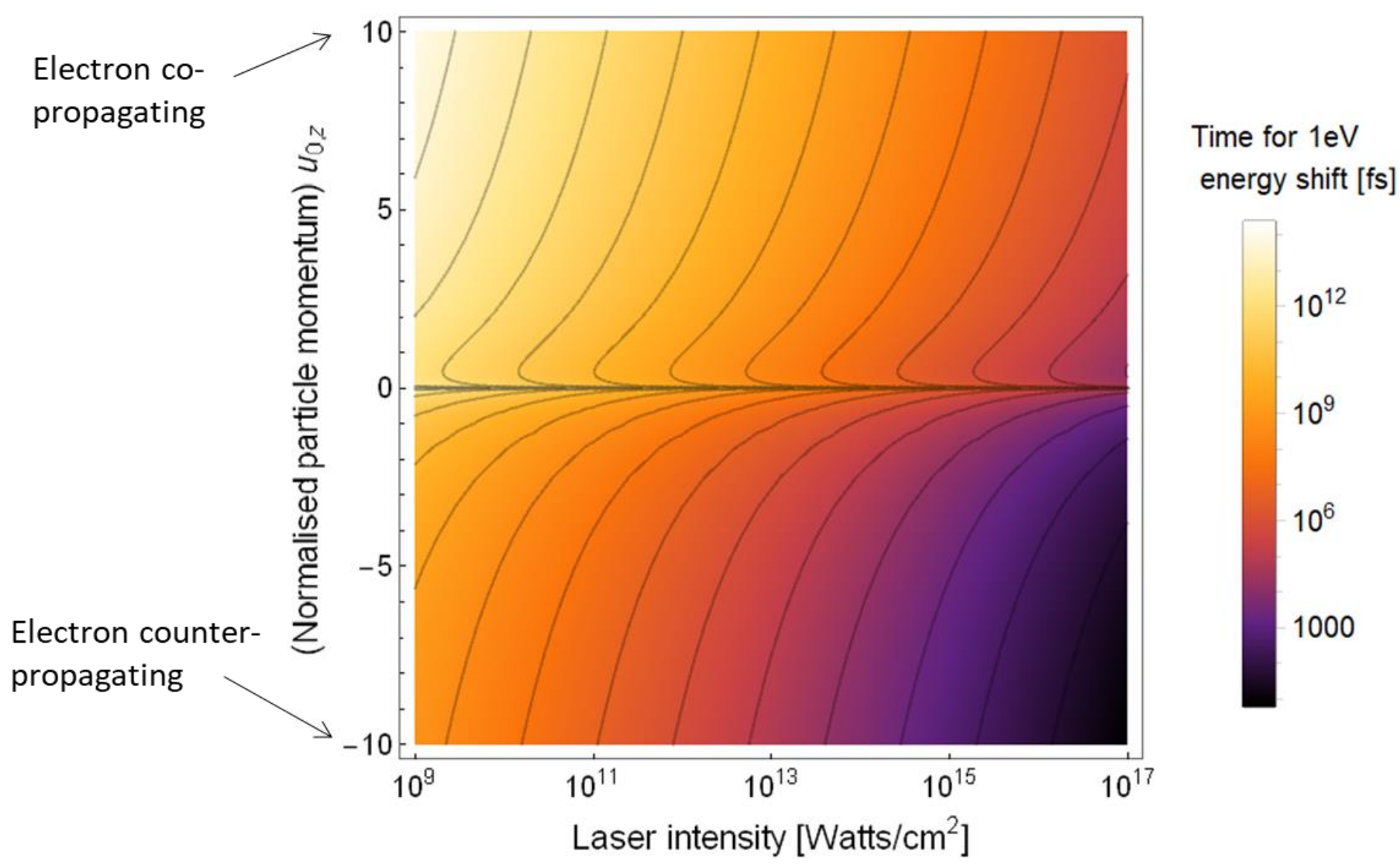

Figure 5: **The characteristic time of energy shift for LL interaction**. The time taking the particle to change its energy by 1 eV, for a pulse with wavelength $532\ \mathrm{nm}$. Note that for $u_0 < 0$, the particle loses energy, while for $u_0 > 0$ the particle gains energy.

Rayleigh length that limits the effective interaction length in free space. As a case example, we consider an optical waveguide that is 3 mm long and 1 μm wide (much wider than the typical electron beam width in electron microscopes). The duration of interaction inside the waveguide, for a 200 keV electron ($\beta = 0.7$), is 14 ps. Then the intensity required to obtain $10m$eV of mean energy loss is $8 \times 10^{13}\ \mathrm{Watt/cm^2}$, corresponding to a total pulse energy of 0.4 mJ. This parameters estimate indicates that the regime required for precision corrections of RR is already currently accessible.

We next consider the scale of energy loss or gain that can be measured in current experiments. Electron energy loss spectroscopy (EELS) can measure changes on the scale of sub-eV (with state-of-the-art systems even reaching the single meV range). Such energy resolution is already below the typical energy of a single photon in the EM field, which hints that we should take into consideration effects related to the second quantization of the field, or the quantum nature of the electron-light interaction [54] These considerations imply that the electron radiates in quantized integer portions of $\hbar\omega$ (that could potentially be described by transitions between quantum Volkov states [55]).

In the setup described above, we have $\chi_0 = k \cdot u \frac{E_0}{F_{\mathrm{cr}}} \ll 1$, with $F_{\mathrm{cr}} = \frac{4\pi\varepsilon_0 m^2 c^3}{q\hbar} = 1.3 \times 10^{16}\ \mathrm{V/cm}$ being the QED critical field. In this regime, the RR energy loss is often considered classically [3], and modeled with conventional equations such as LL. We expect, but cannot prove within our analysis, quantum corrections to appear already at the weak field limit of RR that we study here. Specifically, this expectation arises from current experiments in ultrafast transmission electron microscopy, and especially photon-induced nearfield electron microscopy (PINEM) [49], which show a discrete electron energy change corresponding to a integer number of photons emission/absorption. Such experiments hint that precision measurements of RR may show a mean energy loss as predicted by the classical equations, but still have their exact distribution modified by quantum effects [56]. A more thorough analysis could involve Monte-Carlo estimates of the energy loss and is left for future work. Note also that, for the task of discriminating between classical RR models, such as the LL and LAD equations, which are similar in the first order in perturbation theory, one must consider interactions in which the magnitude of the RR term in the equation of motion is similar in magnitude to the LF term, as was done in [17, 16].

Other experimental setups can also be set to explore these phenomena. For example, we can achieve an interaction over a longer duration ("infinitely-long") by an electron trap or a storage ring under the illumination of a stable continuous wave laser, or a longer duration phase-locked pulse. Going beyond the scheme analyzed in this work, we can consider interactions with two counter-propagating plane waves, or more generally, specially-shaped beams and pulses that will be optimized to provide the maximal RR correction.

In precision measurements of RR, it will be crucial to distinguish the intrinsic RR effects from other effects arising from the deviation of the pulse profile from an ideal plane wave. Two main effects should be considered in this context: the transverse motion of the electron, which might result in the electron escaping from the interaction area, and the ponderomotive acceleration. For the first, we note that the amplitude of transverse oscillation is known to be of order $\lambda a_0$ [57], which is much smaller than the laser spot size for the weak fields we consider here (small $a_0$). To account for the ponderomotive effect, we perform exact simulations of the LL equation in a realistic laser field. The results of our calculations are shown in Figure 6. One can see that RR energy loss can be an order of magnitude stronger than the loss due to ponderomotive scattering. While one cannot "switch-off" the RR force, the two can be distinguished by different scaling laws with respect to the laser energy. Additional details on the simulation can be found in Appendix V.

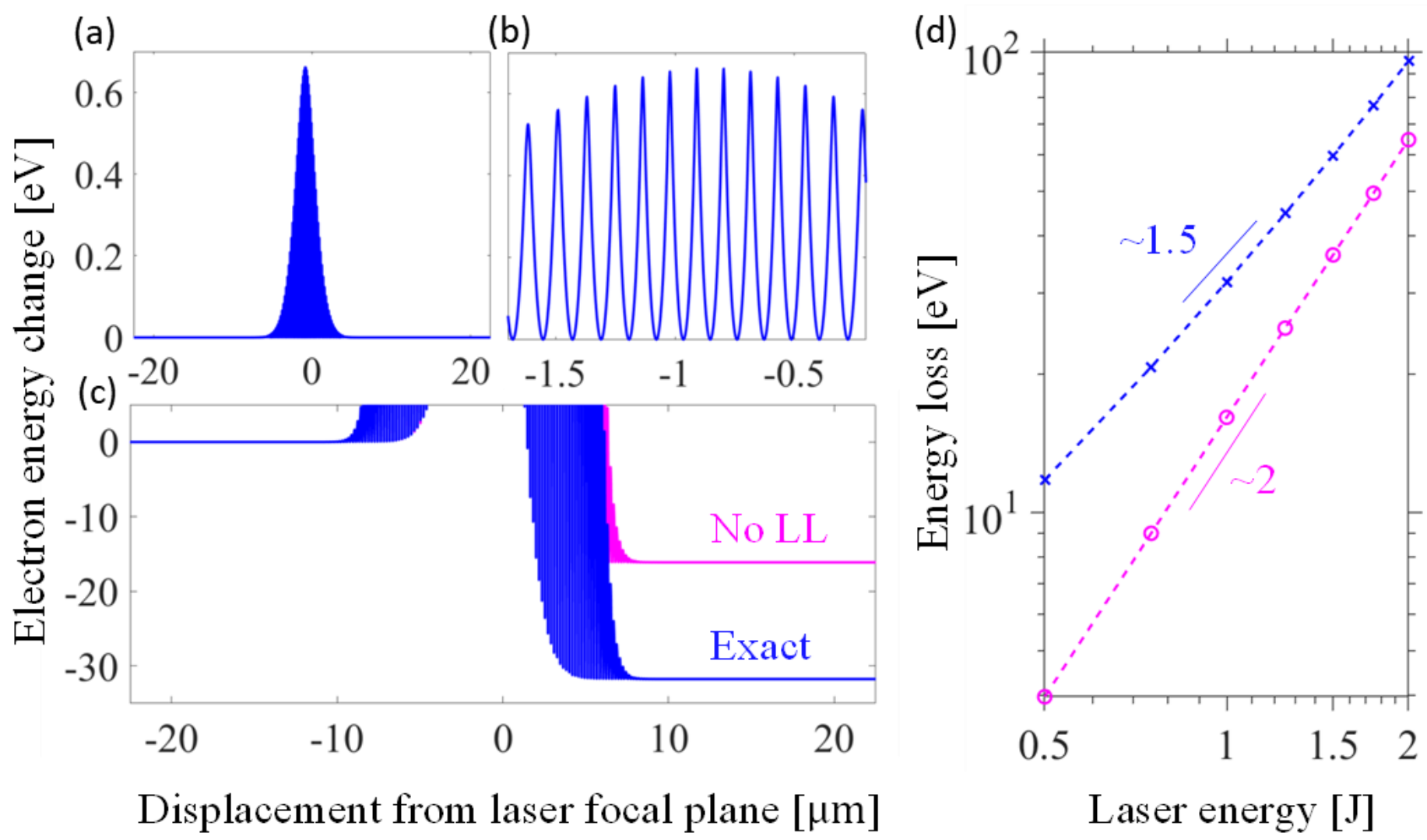

Figure 6: **Simulations of the LL equation in a realistic laser wavepacket.** Results for a 480 keV electron interacting with a counterpropagating laser pulse of duration 30 fs and waist radius 10 µm (peak wavelength 0.8 µm). (a)-(c) show the change in electron energy as a function of the electron displacement along the laser beam axis (for pulse energy of 1 J) compared with the case where the LL terms are removed (no RR). Displacement zero corresponds to the laser's focal plane. (d) shows the very different scaling trends of the electron energy loss between cases with and without radiation reaction.

## Section VII – Conclusions and outlook

Looking forward, the experimental opportunities provided by laser-driven electron microscopy and diffraction (as in the UTEM, DTEM, and UED systems) seem promising for precision measurements of RR. These systems are especially promising for studying the quantized nature of RR in weak EM fields, especially considering the observations of quantized exchange of energy between the electron and the laser field [58, 59], and phenomena like laser-driven quantum walk and Rabi oscillations [46]. Motivated by these observations, it would be fascinating to find similar kinds of underlying physics in the quantum theory of RR.

It is possible that RR effects will also prove to be dependent on the shape of the electron's quantum wavepacket. Recent works suggest the possibilities of controlling the temporal profile of the electron wavepacket [54, 60], showing that certain features of the electron–light interaction can depend on the electron wavepacket [61, 62, 63, 64, 65, 66, 67].

Known effects such as photon-induced nearfield electron microscopy [58] were shown to be wavepacket-dependent [51, 62, 63, 60, 67]. More recent advances have recently unveiled the entanglement between the electron and the radiation it emits [65, 68], which may be a necessary component of the full quantum theory of RR. Using such experiments as an approach to studying the physics of RR will enrich the problem and further enhance the capabilities to measure RR-type effects. Ultimately, we hope that such new capabilities will help probe the inherent quantum nature of RR.

## Appendix

### I. Non-periodicity & divergence of solutions for LAD equation

We show here a proof that there exists no periodic solution of LAD for a particle in a plane wave. To do so, we contract the LAD equation (3) with $k_\alpha$ using equation (6) and obtain

$$k \cdot \dot{u} = \tau_0 (k \cdot \ddot{u} + (\dot{u})^2 \, k \cdot u). \tag{A1.1}$$

Assuming $u$ is a periodic function in $\tau$, then $k \cdot u$ is clearly also a periodic, positive-definite function. We integrate both sides with respect to $\tau$:

$$k \cdot u - \tau_0 k \cdot \dot{u} = \tau_0 \int (\dot{u})^2 (k \cdot u) \, d\tau + C \tag{A1.2}$$

The LHS is periodic, but the RHS is an integral over a non-positive (and non-zero) function, as

$$\dot{u}^\mu = \frac{d}{d\tau}[\gamma(1, \boldsymbol{\beta})] = \left(\gamma^3 (\boldsymbol{\beta} \cdot \dot{\boldsymbol{\beta}}), \gamma^3 (\boldsymbol{\beta} \cdot \dot{\boldsymbol{\beta}}) \boldsymbol{\beta} + \gamma \dot{\boldsymbol{\beta}}\right) \tag{A1.3}$$

$$(\dot{u})^2 = -\gamma^2 \left(\dot{\beta}^2 + \gamma^2 \left(\boldsymbol{\beta} \cdot \dot{\boldsymbol{\beta}}\right)^2\right) \leq 0. \tag{A1.4}$$

The RHS is, therefore, a monotonously decreasing (non-constant, as $\dot{\beta}$ cannot be identically zero in a driving field) function while the LHS is periodic, a contradiction.

We can further use equation (A1.2) to show that for a linearly polarized plane wave, the solution always asymptotically diverges with $k \cdot u \rightarrow 0$. To do so, we note that, because $\xi$ is monotonously increasing, at certain times $\{\tau_i\}$ the particle is in a node of the wave, for which $F^{\mu\nu} = 0$. In that case, the field is locally weak around $\tau_i$ and thus the LF solution with $k \cdot \dot{u} = 0$ is valid. Here we rely on the argument by Spohn et al. [4] showing that the physical solutions of the LAD equation lie on a critical manifold described, for weak fields, by the LL equation (notice that we do not require that LAD generally reduces to LL, but only that it does when the field is weak). Taking the integral limits of (A1.2) between two such instances, we obtain

$$\mathrm{k} \cdot \mathrm{u}(\tau_{\mathrm{i+1}}) - k \cdot u(\tau_i) = \tau_0 \int_{\tau_i}^{\tau_{i+1}} (\dot{u})^2 (k \cdot u) d\tau \,. \tag{A1.5}$$

Because the RHS is negative, we see that $k \cdot u(\tau_i)$ are monotonously decreasing. To show that $k \cdot u(\tau_i) \rightarrow 0$, we assume by contradiction that $k \cdot u(\tau_i) \rightarrow L > 0$. In this case, near $\tau_i$, where the field is weak such that $\dot{u}^\mu \approx \frac{e}{m} F^{\mu\nu} u_\nu$, and for large enough $i$, we have

$$\begin{aligned}
(\dot{u})^2 &\approx \left( \frac{e^2}{m^2} F^{\mu\nu} u_\nu F_{\mu\sigma} u^\sigma \right) \\
&= \left( \frac{e^2}{m^2} (k^\mu A^\nu - k^\nu A^\mu) u_\nu \left( k_\mu A_\sigma - k_\sigma A_\mu \right) u^\sigma \right) \\
&= \frac{e^2 A^\mu A_\mu}{m^2} (k \cdot u)^2 \approx \frac{L^2 e^2 A^\mu A_\mu}{m^2} .
\end{aligned} \tag{A1.6}$$

Where $A^\mu$ is a 4-polarisation vector of the wave satisfying $A^2 = -\left( \frac{E_0}{\omega} \varphi(\xi) \right)^2$ and we used $k \cdot A = k \cdot k = 0$ [15]. Assuming that for some $\epsilon$ range of $\xi(\tau_i)$ the field can be regarded as weak, the integral in equation (A1.5) can be bounded as (denoting $\xi(\tau_i) = \xi_i$ and taking $\varphi(\xi) \approx \xi - \xi_i$ around $\tau_i$)

$$|\mathrm{k} \cdot \mathrm{u}(\tau_{\mathrm{i+1}}) - k \cdot u(\tau_i)| > \left| \tau_0 \int_{\tau_i}^{\tau(\xi = \xi_i + \epsilon)} (\dot{u})^2 (k \cdot u) \, d\tau \right|$$

$$= \left| \tau_0 \int_{\tau_i}^{\tau(\xi=\xi_i+\epsilon)} (\dot{u})^2 \frac{d\xi}{d\tau}\, d\tau \right|$$

$$= \left| \tau_0 \int_{\xi_i}^{\xi_i+\epsilon} (\dot{u})^2 d\xi \right| \approx \tau_0 \frac{\epsilon^2}{2} \frac{L^2 e^2 E_0^2}{m^2 \omega^2}. \qquad \text{(A1.7)}$$

Therefore, the decrease of $k \cdot u(\tau_i)$ in each period is bounded by a finite non-zero amount, in contradiction with the assumption of a finite limit. We have thus shown that in the context of the LAD the velocity will always converge to the speed of light, with the particle traveling along the direction of the wave. In fact, this result shows that the LL dynamics are an attractor of the LAD solution for all initial conditions. The attractor property can be understood as follows: as the particle velocity tends to $c$ in the direction of the wave, both the frequency and field strength go to zero in the frame of reference moving with the particle's average velocity. When the rest-frame field becomes weak, the particle enters a regime in which the LL equation is known to be justified [3].

## II. First-order perturbation for a general radiation force

We develop a first-order perturbation solution for:

$$\dot{u}^\mu = \frac{q}{m} F^\mu_{\ \nu} u^\nu + \varepsilon D^\mu, \qquad \text{(A2.1)}$$

where $F^\mu_{\ \nu}$ is an EM wave field tensor, and $D^\mu\left(u, \dot{u}, \ldots, F^{\mu\nu}, F^{\mu\nu}_{\ \ ,\rho}, \ldots\right)$ is an arbitrary radiation damping term, and ε is a parameter with a dimension of time, so that $\varepsilon\omega_0 \ll 1$. We expand both $u^\mu$ and the differential operator $d_\tau$ in a power series, as

$$\frac{d}{d\tau} = \frac{d\xi}{d\tau}\frac{d}{d\xi} = \left(k \cdot u_{\text{in}} + \varepsilon \dot{\xi}^{(1)} + O(\varepsilon^2)\right) \frac{d}{d\xi} \qquad \text{(A2.2)}$$

$$u^\mu = u^{(0)\mu}(\xi) + \varepsilon u^{(1)\mu}(\xi) + O(\varepsilon^2). \qquad \text{(A2.3)}$$

Substituting in equation (A2.1) and making a change of variables to $\xi$, we obtain the zeroth-order solution, similar to the LF solution

$$u^{(0)\mu}(\xi) = U^{\mu}{}_{\nu} u^{\nu}_{\text{in}}, \tag{A2.4}$$

with $u^{\mu}_{in}$ being the initial velocity and $U^{\mu}{}_{\nu}$ as shown in equation (5). Our interest, however, is in the first order RR correction. In the first-order equation (A2.1) is

$$(k\cdot u_{\text{in}})\frac{d}{d\xi}u^{(1)\mu} + \dot{\xi}^{(1)}\frac{d}{d\xi}u^{(0)\mu} = \frac{q}{m}F^{\mu}_{\ \nu}u^{(1)\nu} + D^{(0)\mu}\,, \tag{A2.5}$$

with

$$D^{(0)\mu} = D^{\mu}\left(u^{(0)}, \dot{u}^{(0)}, F^{\mu\nu}, F^{\mu\nu}_{\ \ ,\rho}\dots\right) = D^{\mu}\left(U^{\mu}{}_{\nu}u^{\nu}_{\text{in}}, \frac{q}{m}F^{\mu}_{\ \nu}U^{\nu}{}_{\sigma}u^{\sigma}_{\text{in}}, F^{\mu\nu}, F^{\mu\nu}_{\ \ ,\rho}\dots\right). \tag{A2.6}$$

Contracting with $k_{\mu}$ and using the identities

$$\dot{\xi}^{(1)} = k\cdot u^{(1)} \tag{A2.7}$$

$$\frac{d}{d\xi}\left(k\cdot u^{(0)}\right) = 0 \tag{A2.8}$$

$$k_{\mu}F^{\mu}{}_{\nu} = 0, \tag{A2.9}$$

we obtain

$$\dot{\xi}^{(1)}(\xi) = \frac{1}{k\cdot u_{\text{in}}}\int_0^{\xi} k\cdot D^{(0)}d\tilde{\xi}\,. \tag{A2.10}$$

Note that in the main paper, e.g. equation (18), $\dot{\xi}^{(1)}$ will be denoted by the auxiliary function $X$. Equation (A2.5) is now a linear ODE of $u^{(1)\mu}$. The homogeneous solution is already known (similar to equation (A2.4)) and we use variation of parameters to find the particular solution. Writing $u^{(1)\mu} = U^{\mu}{}_{\nu}v^{\nu}$, we obtain

$$(k\cdot u_{\text{in}})U^{\mu}{}_{\nu}\frac{d}{d\xi}\,v^{\nu} + \dot{\xi}^{(1)}\frac{d}{d\xi}u^{(0)\mu} = D^{(0)\mu} \tag{A2.11}$$

which, upon integration, yields

$$u^{(1)\mu} = \frac{1}{k\cdot u_{\text{in}}}U^{\mu}{}_{\nu}\int_0^{\xi}(U^{-1})^{\nu}{}_{\sigma}\left[D^{(0)\sigma} - \dot{\xi}^{(1)}\cdot\frac{d}{d\tilde{\xi}}u^{(0)\sigma}\right]d\tilde{\xi}\,. \tag{A2.12}$$

## III. The Hartemann-Luhmann (HL) model for RR

Following the derivation due to Hartemann-Luhmann, the most intuitive way to write the radiation damping force (3-vector) is as an integral over the solid angle of the radiation per unit time

$$\boldsymbol{F} = -\frac{q^2}{16\pi^2\varepsilon_0 c^2}\oint \boldsymbol{n}\frac{\left\{\boldsymbol{n}\times\left[(\boldsymbol{n}-\boldsymbol{\beta})\times\dot{\boldsymbol{\beta}}\right]\right\}^2}{(1-\boldsymbol{\beta}\cdot\boldsymbol{n})^5}d\Omega\,. \tag{A3.1}$$

The integral was then carried out (see [8]) in a frame in which the velocity and acceleration are collinear, yielding

$$\boldsymbol{F} = -m\tau_0\boldsymbol{\beta}\dot{\beta}^2\gamma^6, \tag{A3.2}$$

which was then generalized to a covariant term of the form

$$\begin{aligned} f^{\alpha} &= m\tau_0\dot{u}^2u^{\alpha} \\ &= -\gamma\frac{q^2}{16\pi^2\varepsilon_0 c^2}\oint\frac{\left\{\boldsymbol{n}\times\left[(\boldsymbol{n}-\boldsymbol{\beta})\times\dot{\boldsymbol{\beta}}\right]\right\}^2}{(1-\boldsymbol{\beta}\cdot\boldsymbol{n})^5}(1,\boldsymbol{n})d\Omega\,. \end{aligned} \tag{A3.3}$$

The resulting term matches the nonlinear term in the LAD model presented in equation (3) but lacks the $\ddot{u}^{\mu}$ term. When added to the LF, the resulting equation does not preserve the norm $u\cdot u = 1$, which is required by the definition of the 4-velocity, as it does not satisfy

$$f\cdot u = 0. \tag{A3.4}$$

Naively, one might try to "fix" equation (A3.1) by keeping the space components of the 4-force but replacing the time component so that equation (A3.4) will be satisfied. This procedure gives the following expression for the RR term

$$f^{\alpha} = -\gamma\frac{q^2}{16\pi^2\varepsilon_0 c^2}\oint\frac{\left\{\boldsymbol{n}\times\left[(\boldsymbol{n}-\boldsymbol{\beta})\times\dot{\boldsymbol{\beta}}\right]\right\}^2}{(1-\boldsymbol{\beta}\cdot\boldsymbol{n})^5}(\boldsymbol{n}\cdot\boldsymbol{\beta},\boldsymbol{n})d\Omega\,. \tag{A3.5}$$

This expression can be plugged into equation (20) to obtain the phase space shown in Figure 3b. This space displays a steady-state point, which according to our proof cannot occur for a Lorentz invariant equation of motion. Indeed, the expression (A3.5) is not Lorentz covariant and therefore is not a viable equation of motion.

## IV. Deriving the relation of laser fluence and energy loss due to the LL term (22)

Following the derivation in [15], we can write the following equation for the energy change as a function of $\xi$

$$s=\frac{\Delta E(\xi)}{mc^2}=\frac{a_0^2\tau_0\omega_0\xi\left(2u_{\text{in}}^2+1-2u_{\text{in}}\sqrt{1+u_{\text{in}}^2}\right)\left(2u_{\text{in}}+\frac{1}{2}a_0^2\tau_0\omega_0\xi\right)}{4\left(\sqrt{1+u_{\text{in}}^2}-u_{\text{in}}\right)\left(1+\frac{1}{2}a_0^2\tau_0\omega_0\xi\left(\sqrt{1+u_{\text{in}}^2}-u_{\text{in}}\right)\right)}, \tag{A4.1}$$

where $u_{\text{in}}$ is the 4-velocity component in the direction of the wave vector ($u_{\text{in}}>0$ for a copropagating electron). We can solve for $\xi$ and obtain an expression for the pulse duration $T=\frac{\xi}{\omega_0}$. A simpler expression will be obtained by noting that $\beta_{\text{in}}=\frac{u_{\text{in}}}{\sqrt{1+u_{\text{in}}^2}}=\frac{u_{\text{in}}}{\gamma_{\text{in}}}$. We then get

$$T=2\frac{\gamma_{\text{in}}\beta_{\text{in}}+s\mp\gamma_{\text{in}}\sqrt{\beta_{\text{in}}^2+s\left(\frac{s}{\gamma_{\text{in}}^2}+\frac{2}{\gamma_{\text{in}}}\right)}}{a_0^2\tau_0\omega_0^2}, \tag{A4.2}$$

(where we have a minus sign for $\beta_{\text{in}}>0$, plus for $\beta_{\text{in}}<0$). Assuming $s\ll\beta^2$ (as can be true for electron energy change of less than 1 MeV and weakly relativistic initial velocity), we can expand equation (A4.2) as

$$T=\frac{2s}{a_0^2\tau_0\omega_0^2}\left(1+\frac{1}{\beta_{\text{in}}}\right). \tag{A4.3}$$

The wave intensity is then given by

$$I=\frac{\varepsilon_0cE^2}{2}=\frac{a_0^2\omega_0^2c^3m^2}{2q^2}, \tag{A4.4}$$

so that the total fluence is

$$H=I\cdot T=\frac{cm_e\varepsilon_0\Delta E}{q^2\tau_0}\left(1+\frac{1}{\beta_{\text{in}}}\right). \tag{A4.5}$$

The reader can see that if the fluence is known, the energy change is independent of the specific laser parameters $\omega_0, a_0$ and $T$. More generally, it is also independent of the shape of the pulse envelope.

## V. Simulations in a realistic laser pulse field

To account for effects arising from the presence of a realistic laser field, we performed exact simulations of the LL equations in a counterpropagating laser pulse. In the simulations we consider a sub-relativistic electron of kinetic energy 480 keV (obtained, for instance, from a transmission electron microscope) interacting with a counterpropagating 1 J laser pulse. The results are obtained using the *ab initio* software described in the Supplementary Information of [35]. The laser pulse is modeled using an exact, fully closed form, nonparaxial solution to Maxwell's equations [69]. In Fig. 6, we note that the net change in electron energy with and without radiation reaction can be readily resolved by typical electron energy loss spectroscopy (EELS) setups [70]. Furthermore, Fig. 6d shows the distinctive trend (that scales as the power of 1.5 as opposed to 2) that the energy loss spectrum would exhibit if RR, as described by the LL equation, is present. We note that our parameters were chosen to ensure that the electron energy loss is well within typical EELS resolutions and that the contribution of RR is at least comparable with the contribution of ponderomotive deceleration. These parameters can be relaxed quite significantly, allowing for even lower laser pulse energies and electron pulse energies, with sufficiently sensitive EELS setups.

## Bibliography

[1] J. D. Jackson, Classical Electromagnetism, 2nd ed., American Journal of Physics, 1999.

[2] P. A. M. Dirac, "Classical Theory of Radiating Electrons," *Proc. R. Soc. A,* pp. 148-169, 1938.

[3] A. Di Piazza, C. Müller, K. Z. Hatsagortsyan and C. H. Keitel, "Extremely high-intensity laser interactions with fundamental quantum systems," *Rev. Mod. Phys.,* vol. 84, no. 3, pp. 1177-1228, 2012.

[4] H. Spohn, "The critical manifold of the Lorentz-Dirac equation," *Europhys. Lett.,* vol. 50, no. 3, pp. 287-292, 2000.

[5] C. J. Eliezer, "On the classical theory of particles," *Proc. R. Soc. A,* vol. 194, no. 1039, pp. 543-555, 1948.

[6] L. Landau and E. M. Lifshitz, The Classical Theory of Fields, American Journal of Physics 21, 1953.

[7] T. C. Mo and C. H. Papas, "New Equation of Motion for Classical Charged Particles," *PRD,* vol. 4, no. 12, pp. 3566-3571, 1971.

[8] F. V. Hartemann and N. C. Luhmann, "Classical electrodynamical derivation of the radiation damping force," *PRL,* vol. 74, no. 7, pp. 1107-1110, 1995.

[9] I. V. Sokolov, "Renormalization in Lorentz-Abraham-Dirac Equation, Describing Radiation Force in Classical Electrodynamics," *arXiv:0906.1150,* 2009.

[10] P. Caldirola, "A relativistic theory of the classical electron," *Riv. Nuovo Cimento Soc. Ital. Fis.,* vol. 2, no. 13, pp. 1-49, 1979.

[11] C. Bild, D. A. Deckert and H. Ruhl, "Radiation reaction in classical electrodynamics," *Phys. Rev. D,* vol. 99, no. 096001, 2019.

[12] Y. Yaremko, "Exact solution to the Landau-Lifshitz equation in a constant electromagnetic field," *J. Math. Phys.,* vol. 54, no. 9, p. 092901, 2013.

[13] R. Rivera and D. Villarroel, "Exact solutions to the Mo-Papas and Landau-Lifshitz equations," *PRE,* vol. 4, no. 66, p. 046618, 2002.

[14] A. Di Piazza, "Exact solution of the Landau-Lifshitz equation in a plane wave," *Lett. Math. Phys.,* vol. 83, no. 3, pp. 305-313, 2008.

[15] Y. Hadad, L. Labun, J. Rafelski, N. Elkina, C. Klier and H. Ruhl, "Effects of radiation reaction in relativistic laser acceleration," *PRD,* vol. 82, no. 9, 2010.

[16] J. M. Cole, K. T. Behm, E. Gerstmayr, T. G. Blackburn, J. C. Wood, C. D. Baird, M. J. Duff, C. Harvey, A. Ilderton, A. S. Joglekar, K. Krushelnick, S. Kuschel, M. Marklund, P. McKenna, C. D. Murphy, K. Poder, C. P. Ridgers and G. M. Samarin, "Experimental evidence of radiation reaction in the collision of a high-intensity laser pulse with a laser-wakefield accelerated electron beam," *PRX,* vol. 8, no. 3, 2018.

[17] K. Poder, M. Tamburini, G. Sarri, A. Di Piazza, S. Kuschel, C. D. Baird, K. Behm, S. Bohlen, J. M. Cole, D. J. Corvan, M. Duff, E. Gerstmayr, C. H. Keitel, K. Krushelnick, S. P. Mangles, P. McKenna, C. D. Murphy, Z. Najmudin and Ridgers, "Experimental Signatures of the Quantum Nature of Radiation Reaction in the Field of an Ultraintense Laser," *PRX,* vol. 8, no. 3, p. 031004, 2018.

[18] T. N. Wistisen, A. Di Piazza, H. V. Knudsen and U. I. Uggerhøj, "Experimental evidence of quantum radiation reaction in aligned crystals," *Nature Comm.,* vol. 9, no. 1, p. 795, 2018.

[19] T. Heinzl, C. Harvey, A. Ilderton, M. Marklund, S. S. Bulanov, S. Rykovanov, C. B. Schroeder, E. Esarey and W. P. Leemans, "Detecting radiation reaction at moderate laser intensities," *PRE,* vol. 91, p. 023207, 2015.

[20] F. V. Hartemann, S. N. Fochs, G. P. Le Sage, N. C. Luhmann, J. G. Woodworth, M. D. Perry, Y. J. Chen and A. K. Kerman, "Nonlinear pondermotive scattering of relativistic electrons by an intense laser field at focus," *PRE,* vol. 51, no. 5, pp. 4833-4848, 1995.

[21] C. Harvey, T. Heinzl and M. Marklund, "Symmetry breaking from radiation reaction in ultra-intense laser fields," *PRD,* vol. 84, no. 11, p. 116005, 2011.

[22] P. O. Kazinski, "Radiation of de-excited electrons at large times in a strong," *Annals Phys.,* vol. 339, pp. 430-459, 2013.

[23] S. R. Yoffe, Y. Kravets, A. Noble and D. A. Jaroszynski, "Longitudinal and transverse cooling of relativistic electron beams," *New J. Phys.,* vol. 17, no. 5, p. 053025, 2015.

[24] V. Dinu, C. Harvey, A. Ilderton, M. Marklund and G. Torgrimsson, "Quantum radiation reaction: from interference to incoherence," *Phys. Rev. Lett,* vol. 116, no. 4, p. 044801, 2016.

[25] R. W. Nelson and I. Wasserman, "Synchrotron radiation with radiation reaction," *Astrophysical Journal,* vol. 371, pp. 265-276, 1991.

[26] L. Schächter, Beam-Wave Interaction in Periodic and Quasi-Periodic Structures, Berlin, Heidelberg: Springer, 2011.

[27] V. I. Veksler, "The principle of coherent acceleration of charged particles," *The Soviet Journal of Atomic Energy,* vol. 2, pp. 525-528, 1957.

[28] E. J. Moniz and D. H. Sharp, "Radiation Reaction in Nonrelativistic Quantum Electrodynamics," *Phys. Rev. D,* vol. 15, p. 2850, 1977.

[29] V. S. Krivitsky and V. N. Tsytovich, "Average radiation reaction force in quantum electrodynamics," *Sov. Phys. Usp.,* vol. 34, pp. 250-258, 1991.

[30] A. Higuchi, "Radiation reaction in quantum field theory," *Phys. Rev. D,* vol. 66, p. 105004, 2002.

[31] A. Ilderton and G. Torgrimsson, "Radiation reaction in strong field QED," *Phys. Lett. B,* vol. 725, p. 481, 2013.

[32] A. Gonoskov, T. G. Blackburn, M. Marklund and S. S. Bulanov, "Charged particle motion and radiation in strong electromagnetic fields," *arXiv,* p. 2107.02161, 2021.

[33] J. D. Lawson, "Lasers and accelerators," *IEEE Trans. Nuc. Sci.,* vol. 26, no. 3, pp. 4217-4219, 1979.

[34] P. Woodward, "A method of calculating the field over a plane aperture required to produce a given polar diagram," *Journal of the Institution of Electrical Engineers - Part IIIA: Radiolocation,* vol. 93, no. 10, pp. 1554-1558, 1946.

[35] L. J. Wong, "Laser-Induced Linear-Field Particle Acceleration in Free Space," *Sci. Rep.,* vol. 7, no. 1, p. 11159, 2017.

[36] M. Tamburini, C. H. Keitel and A. D. Piazza, "Electron dynamics controlled via self-interaction," *PRE,* vol. 89, no. 2, p. 021201, 2014.

[37] A. Prosperetti, "The motion of a charged particle in a uniform magnetic field," *A. Nuov. Cim. B,* vol. 57, no. 2, pp. 253-268, 1980.

[38] L. S. Brown and T. W. B. Kibble, "Interaction of intense laser beams with electrons," *Phys. Rev.,* vol. 133, no. 3A, pp. A705-A719, 1964.

[39] "With the possible exceptions of the Caldirola and Sokolov equations, also see [33]".

[40] S. H. Strogatz, Nonlinear dynamics and chaos: with applications to physics, biology, chemistry, and engineering, 2nd ed., Boulder, Co: Westview Press, 2015.

[41] R. T. Hammond, "Relativistic particle motion and radiation reaction in electrodynamics," *EJTP,* vol. 6, no. 23, pp. 221-258, 2010.

[42] V. B. Berestetskii, L. P. Pitaevskii and E. M. Lifshitz, Quantum Electrodynamics, 2nd ed., Butterworth-Heinemann, 1982.

[43] "The inclusion of only the \kappa^2 terms in equation (23) is necessary for the proof. Otherwise different terms of g_n can cancel each other only for a certain choice of kappa.".

[44] I. V. Sokolov, N. M. Naumova, J. A. Nees, Ǵ. A. Mourou and V. P. Yanovsky, "Dynamics of emitting electrons in strong laser fields," *Phys. Plasmas,* vol. 16, no. 9, p. 093115, 2009.

[45] T. LaGrange, G. Campbell, B. Reed, M. Taheri, J. Pesavento, J. Kim and N. Browning, "Nanosecond time-resolved investigations using the in situ of dynamic transmission electron microscope (DTEM)," *Ultramicroscopy,* vol. 108, no. 11, pp. 1441-1449, 2008.

[46] A. Feist, K. E. Echternkamp, J. Schauss, S. V. Yalunin, S. Schäfer and C. Ropers, "Quantum coherent optical phase modulation in an ultrafast transmission electron microscope," *Nature,* vol. 521, pp. 200-203, 2015.

[47] T. LaGrange, B. W. Reed and D. J. Masiel, "Movie-mode dynamic electron microscopy," *MRS Bulletin,* vol. 40, no. 1, pp. 22-28, 2015.

[48] Y. Morimoto and P. Baum, "Diffraction and microscopy with attosecond electron pulse trains," *Nature Phys.,* vol. 14, no. 3, pp. 252-256, 2018.

[49] C. Kealhofer, W. Schneider, D. Ehberger, A. Ryabov, F. Krausz and P. Baum, "All-optical control and metrology of electron pulses," *Science,* vol. 352, no. 6284, pp. 429-433, 2016.

[50] E. Pomarico, I. Madan, G. Berruto, G. M. Vanacore, K. Wang, I. G. d. A. F. J. Kaminer and F. Carbone, "meV Resolution in Laser-Assisted Energy-Filtered Transmission Electron Microscopy," *ACS Photonics,* vol. 5, no. 3, pp. 759-764, 2018.

[51] K. E. Priebe, C. Rathje, S. V. Yalunin, T. Hohage, A. Feist, S. Schäfer and C. Ropers, "Attosecond electron pulse trains and quantum state reconstruction in ultrafast transmission electron microscopy," *Nature Phot.,* vol. 11, no. 12, pp. 793-797, 2017.

[52] A. Mizrahi and L. Schächter, "Optical Bragg accelerators," *PRE,* vol. 70, no. 1, p. 016505, 2004.

[53] K. J. Leedle, R. Fabian Pease, R. L. Byer and J. S. Harris, "Laser acceleration and deflection of 96.3 keV electrons with a silicon dielectric structure," *Optica,* vol. 2, no. 2, p. 158, 2015.

[54] I. Kaminer, J. Nemirovsky, M. Rechtsman, R. Bekenstein and M. Segev, "Self-Accelerating Dirac particles and prolonging the lifetime of relativistic fermions," *Nature Phys.,* vol. 11, no. 3, pp. 261-267, 2015.

[55] A. Di Piazza, K. Z. Hatsagortsyan and C. H. Keitel, "Quantum radiation reaction effects in multiphoton compton scattering," *PRL,* vol. 105, no. 22, p. 220403, 2010.

[56] F. J. García de Abajo, "Optical excitations in electron microscopy," *Rev. Mod. Phys,* vol. 82, no. 1, pp. 209-275, 2010.

[57] N. Rivera, L. J. Wong, M. Soljačić and I. Kaminer, "Ultrafast Multiharmonic Plasmon Generation by Optically Dressed Electrons," *Physical Review Letters,* vol. 122, no. 5, p. 053901, 2019.

[58] B. Barwick, D. J. Flannigan and A. H. Zewail, "Photon-induced near-field electron microscopy," *Nature,* vol. 462, pp. 902-906, 2009.

[59] L. Piazza, T. T. Lummen, E. Quiñonez, Y. Murooka, B. W. Reed, B. Barwick and F. Carbone, "Simultaneous observation of the quantization and the interference pattern of a plasmonic near-field," *Nature Comm.,* vol. 6, no. 1, p. 6407, 2015.

[60] R. Dahan, S. Nehemia, M. Shentcis, O. Reinhardt, Y. Adiv, X. Shi, O. Be'er, M. Lynch, Y. Kurman, K. Wang and I. Kaminer, "Resonant phase-matching between a light

wave and a free-electron wavefunction," *Nature Phys.,* vol. 16, no. 11, pp. 1123-1131, 2020.

[61] L. J. Wong, N. Rivera, C. Murdia, T. Christensen, J. D. Joannopoulos, M. Soljačić and I. Kaminer, "Control of quantum electrodynamical processes by shaping electron wavepackets," *Nature comm.,* pp. 12(1), 1-10, 2021.

[62] G. M. Vanacore, I. Madan, G. Berruto, K. Wang, E. Pomarico, R. J. Lamb, D. McGrouther and I. Kaminer, "From attosecond to zeptosecond coherent control of free-electron wave functions using semi-infinite light fields," *Nature Comm.,* vol. 9, no. 1, p. 2694, 2018.

[63] Y. Pan, B. Zhang and A. Gover, "Anomalous Photon-Induced Near-Field Electron Microscopy," *PRL,* vol. 122, no. 18, p. 183204, 2019.

[64] A. Gover and Y. Pan, "Dimension-dependent stimulated radiative interaction of a single electron quantum wavepacket," *Phys. Lett. A,* vol. 382, no. 23, pp. 1550-1555, 2018.

[65] A. Karnieli, N. Rivera, A. Arie and I. Kaminer, "The coherence of light is fundamentally tied to the quantum coherence of the emitting particle," *Science Advances,* vol. 7, no. 18, p. eabf8096, 2021.

[66] O. Kfir, V. Di Giulio, F. de Abajo and C. Ropers, "Optical coherence transfer mediated by free electrons," *Sicence Advances,* vol. 7, no. 18, p. eabf6380, 2021.

[67] R. Dahan, A. Gorlach, U. Haeusler, A. Karnieli, O. Eyal, P. Yousefi, M. Segev, A. Arie, G. Eisenstein, P. Hommelhoff and I. Kaminer, "Imprinting the quantum statistics of photons on free electrons," *Science,* vol. 373, no. 6561, pp. 1309-1310, 2021.

[68] Y. Adiv, H. Hu, S. Tsesses, R. Dahan, K. Wang, Y. Kurman, A. Gorlach, H. Chen, X. Lin, G. Bartal and I. Kaminer, "Observation of 2D Cherenkov radiation," *arXiv,* p. 2203.01698, 2022.

[69] L. J. Wong, F. X. Kärtner and S. G. Johnson, "Improved beam waist formula for ultrashort, tightly focused linearly, radially, and azimuthally polarized laser pulses in free space," *Optics Letters,* vol. 39, no. 5, pp. 1258-1261, 2014.

[70] R. F. Egerton, Electron energy-loss spectroscopy in the electron microscope, Springer Science & Business Media, 2011.